\documentclass[
  reprint,
  superscriptaddress,
  aps, prl,
  floatfix,
  nobibnotes
]{revtex4-2}

\usepackage{graphicx}
\usepackage{dcolumn}
\usepackage{amsmath}

\usepackage{upgreek}
\usepackage{tikz}

\begin{document}

\title{Disparate Quantum Corrections to Conduction in Carbon Nanotube Bundles}

\author{Shengjie Yu}
\affiliation{Department of Electrical and Computer Engineering, Rice University, Houston, Texas 77005, USA}
\affiliation{Applied Physics Graduate Program, Smalley-Curl Institute, Rice University, Houston, Texas 77005, USA}
\affiliation{Carbon Hub, Rice University, Houston, Texas 77005, USA}

\author{Zhengyi Lu}
\affiliation{Department of Electrical and Computer Engineering, Rice University, Houston, Texas 77005, USA}
\affiliation{Department of Physics and Astronomy, Rice University, Houston, Texas 77005, USA}

\author{Renjie Luo}
\affiliation{Department of Physics and Astronomy, Rice University, Houston, Texas 77005, USA}

\author{Tanner Legvold}
\thanks{Deceased}
\affiliation{Department of Physics and Astronomy, Rice University, Houston, Texas 77005, USA}

\author{Natsumi Komatsu}
\affiliation{Department of Electrical and Computer Engineering, Rice University, Houston, Texas 77005, USA}
\affiliation{Carbon Hub, Rice University, Houston, Texas 77005, USA}
\affiliation{Department of Bioengineering, University of Illinois Urbana–Champaign, Urbana, Illinois 61801, USA}

\author{Liyang~Chen}
\affiliation{Applied Physics Graduate Program, Smalley-Curl Institute, Rice University, Houston, Texas 77005, USA}
\affiliation{Department of Physics and Astronomy, Rice University, Houston, Texas 77005, USA}

\author{Oliver S.\ Dewey}
\affiliation{Carbon Hub, Rice University, Houston, Texas 77005, USA}
\affiliation{Department of Chemical and Biomolecular Engineering, Rice University, Houston, Texas 77005, USA}

\author{Lauren W.\ Taylor}
\affiliation{Carbon Hub, Rice University, Houston, Texas 77005, USA}
\affiliation{Department of Chemical and Biomolecular Engineering, Rice University, Houston, Texas 77005, USA}
\affiliation{Department of Chemical and Biomolecular Engineering, The Ohio State University, Columbus, OH 43210}

\author{Huaijin Sun}
\affiliation{Department of Physics and Astronomy, Rice University, Houston, Texas 77005, USA}

\author{Matteo Pasquali}
\affiliation{Carbon Hub, Rice University, Houston, Texas 77005, USA}
\affiliation{Department of Chemical and Biomolecular Engineering, Rice University, Houston, Texas 77005, USA}
\affiliation{Smalley--Curl Institute, Rice University, Houston, Texas 77005, USA}
\affiliation{Department of Materials Science and NanoEngineering, Rice University, Houston, Texas 77005, USA}
\affiliation{Department of Chemistry, Rice University, Houston, Texas 77005, USA}

\author{Geoff Wehmeyer}
\affiliation{Carbon Hub, Rice University, Houston, Texas 77005, USA}
\affiliation{Smalley--Curl Institute, Rice University, Houston, Texas 77005, USA}
\affiliation{Department of Mechanical Engineering, Rice University, Houston, Texas 77005, USA}

\author{Matthew S.\ Foster}
\affiliation{Carbon Hub, Rice University, Houston, Texas 77005, USA}
\affiliation{Department of Physics and Astronomy, Rice University, Houston, Texas 77005, USA}
\affiliation{Smalley--Curl Institute, Rice University, Houston, Texas 77005, USA}

\author{Junichiro Kono}
\thanks{Corresponding author: kono@rice.edu}
\affiliation{Department of Electrical and Computer Engineering, Rice University, Houston, Texas 77005, USA}
\affiliation{Carbon Hub, Rice University, Houston, Texas 77005, USA}
\affiliation{Department of Physics and Astronomy, Rice University, Houston, Texas 77005, USA}
\affiliation{Smalley--Curl Institute, Rice University, Houston, Texas 77005, USA}
\affiliation{Department of Materials Science and NanoEngineering, Rice University, Houston, Texas 77005, USA}

\author{Douglas Natelson}
\thanks{Corresponding author: natelson@rice.edu}
\affiliation{Department of Electrical and Computer Engineering, Rice University, Houston, Texas 77005, USA}
\affiliation{Carbon Hub, Rice University, Houston, Texas 77005, USA}
\affiliation{Department of Physics and Astronomy, Rice University, Houston, Texas 77005, USA}
\affiliation{Smalley--Curl Institute, Rice University, Houston, Texas 77005, USA} 
\affiliation{Department of Materials Science and NanoEngineering, Rice University, Houston, Texas 77005, USA}

\date{\today}

\begin{abstract}
Quantum interference effects such as weak localization (WL) and universal conductance fluctuations (UCF) normally yield consistent electronic phase-coherence lengths, $L_{\phi}$, in homogeneous conductors.  
Here we show that in individual carbon-nanotube bundles exfoliated from highly conductive solution-spun fibers, different probes—including the field scales and magnitudes of WL and UCF, and nonlocal magnetoconductance—lead to strikingly disparate estimates of coherence lengths.
WL magnetoconductance measured in a perpendicular field gives $L_{\phi} \approx 50$\,nm.  
In contrast, UCF amplitudes are comparable to $e^{2}/h$ even for an 8-$\upmu$m-long segment, and nonlocal magnetoconductance persists across a 4-$\upmu$m separation of electrode, revealing phase-coherent transport over micrometer scales within a single bundle.  
The coexistence of short- and long-range coherence implies that locally diffusive electrons remain partially phase-correlated among nanotubes within the same bundle.
These findings challenge the conventional single-scale picture of mesoscopic coherence and establish carbon nanotube bundles as a model platform for emergent, network-level quantum transport.
\end{abstract}

\maketitle

Quantum interference of diffusive electrons produces reproducible corrections to classical transport—weak localization (WL), universal conductance fluctuations (UCF), and Altshuler–Aronov (AA) interaction effects~\cite{lee1985disordered, altshuler1981, Beenakker1991}.  
In homogeneous metals and semiconductors, analysis of WL magnetoconductance and UCF correlation fields and magnitudes yields consistent estimates of the electronic phase-coherence length, $L_{\phi}$\cite{Aleiner_1999,trionfi2004ucfwl}.  
For carbon nanotubes (CNTs), individual tubes exhibit ballistic transport and quantum interference phenomena such as Fabry–P\'erot resonances and Aharonov–Bohm oscillations~\cite{Bachtold1999, McEuen1999}.  
Macroscopic assemblies of aligned CNTs offer a unique platform to explore quantum coherence in large-scale, flexible conductors.  
However, in such hierarchical systems, the extent to which charge transport between neighboring nanotubes remains phase-coherent—and the role of inelastic scattering in that intertube motion—remains an open question, even though quantum corrections to conduction are highly sensitive to such processes.  
This motivates quantitative comparisons of coherence scales inferred from WL, UCF, and nonlocal magnetotransport within individual CNT bundles.

Dense, highly aligned CNT fibers spun from liquid-crystalline superacid solutions~\cite{Yamaguchi2019, Yu2025} provide such a platform. 
These macroscopic fibers—tens of micrometers in diameter and composed of bundles tens of nanometers across—combine high electrical conductivity with mechanical flexibility and low density.  
Understanding conduction across this hierarchical architecture is crucial for linking quantum interference at the tube and bundle level to macroscopic performance.  
Previous magnetotransport studies on macroscopic CNT fibers have revealed WL- and UCF-like signatures~\cite{baxendale1997magnetotransport, baxendale1998metallic, Yu2025}, but the heterogeneous nature of these systems complicates quantitative extraction and comparison of phase-coherence lengths. 
Isolating and probing individual CNT bundles therefore provides a more controlled setting for examining coherence across multiple transport probes.
However, the extent and dimensionality of phase coherence within individual bundles of macroscopically aligned CNT fibers remain open questions.

Here, we report low-temperature magnetotransport measurements on individual aligned CNT bundles exfoliated from ultrahigh-conductivity solution-spun fibers.  
Surprisingly, we find that detailed comparison of fundamental quantum corrections—the magnetic field dependence of the WL and UCF, and the magnitudes of the UCF and nonlocal magnetoconductance—yield coherence lengths that differ by orders of magnitude.
Specifically, our measurements revealed two distinct regimes of electronic phase coherence coexisting within a single bundle:  
(i)~a short-range ($\sim$50\,nm) diffusive phase coherence associated with WL and the field scale of UCF, arising from narrow flux-enclosing loops with Tesla-scale characteristic fields, and  
(ii)~a micrometer-scale coherent transport manifested by large-amplitude UCFs and nonlocal magnetoconductance that persist up to $\sim$20\,K.
This unexpected dichotomy challenges the conventional assumption of a single characteristic phase-coherence scale.
At the same time, these results show that quantum coherence in aligned CNT bundles extends over micrometer distances and remains robust at elevated temperatures.  
Consequently, assemblies of aligned CNTs provide an effective model system for investigating coherence percolation and emergent quantum connectivity in complex anisotropic conductors.


For low temperature measurements, individual CNT bundles were exfoliated from macroscopic aligned CNT fibers using a dry “Scotch-tape” technique and transferred onto SiO$_2$/Si substrates with pre-patterned alignment marks.  
The parent CNT fibers are spun from raw CNTs that are primarily double-walled, with an average outer-wall diameter of $\sim$1.5\,nm and a viscosity-averaged aspect ratio of $\sim$4000, corresponding to typical tube lengths of order several micrometers (e.g., $\sim$6\,$\upmu$m)~\cite{Yu2025,Tsentalovich2016}.  
The resulting fibers, produced from liquid-crystalline superacid solutions by the solution-spinning method~\cite{Behabtu2013}, exhibit metallic-like temperature dependence of their resistance, with quantum corrections to the conductivity, distinct from the insulating behavior of less ordered CNT materials.  
These commercially available fibers~\cite{TAYLOR2021} possess high alignment and packing density, achieving a record room-temperature conductivity of 10.9\,MS/m among macroscopic CNT materials.  
Residual acid doping in the as-spun fibers ensures a high carrier density, while partial dedoping during device processing of the CNT bundles slightly reduces the conductivity.
Atomic-force microscopy was used to locate candidate bundles, determine their positions relative to the alignment marks, and estimate their diameters (30–80\,nm).  
Multiterminal electrodes were defined by electron-beam lithography followed by Ti/Au ($4/80\,\mathrm{nm}$) evaporation and liftoff, with interlead spacings of 2–8\,$\upmu$m. Multiple segments were fabricated and measured on each bundle.

Transport measurements were performed in Quantum Design PPMS and OptiCool cryostats, with a magnetic field ($B$) up to $\pm$9\,T applied either perpendicular or parallel to the CNT bundle axis.  
Low-frequency lock-in techniques (typical excitation currents of $\sim$20\,nA for local and $\sim$66\,nA for nonlocal measurements) were used to measure the zero-bias differential resistance while minimizing Joule heating and dedoping of the acid-processed CNTs.  

\begin{figure}[t]
  \centering
  \includegraphics[width=\linewidth]{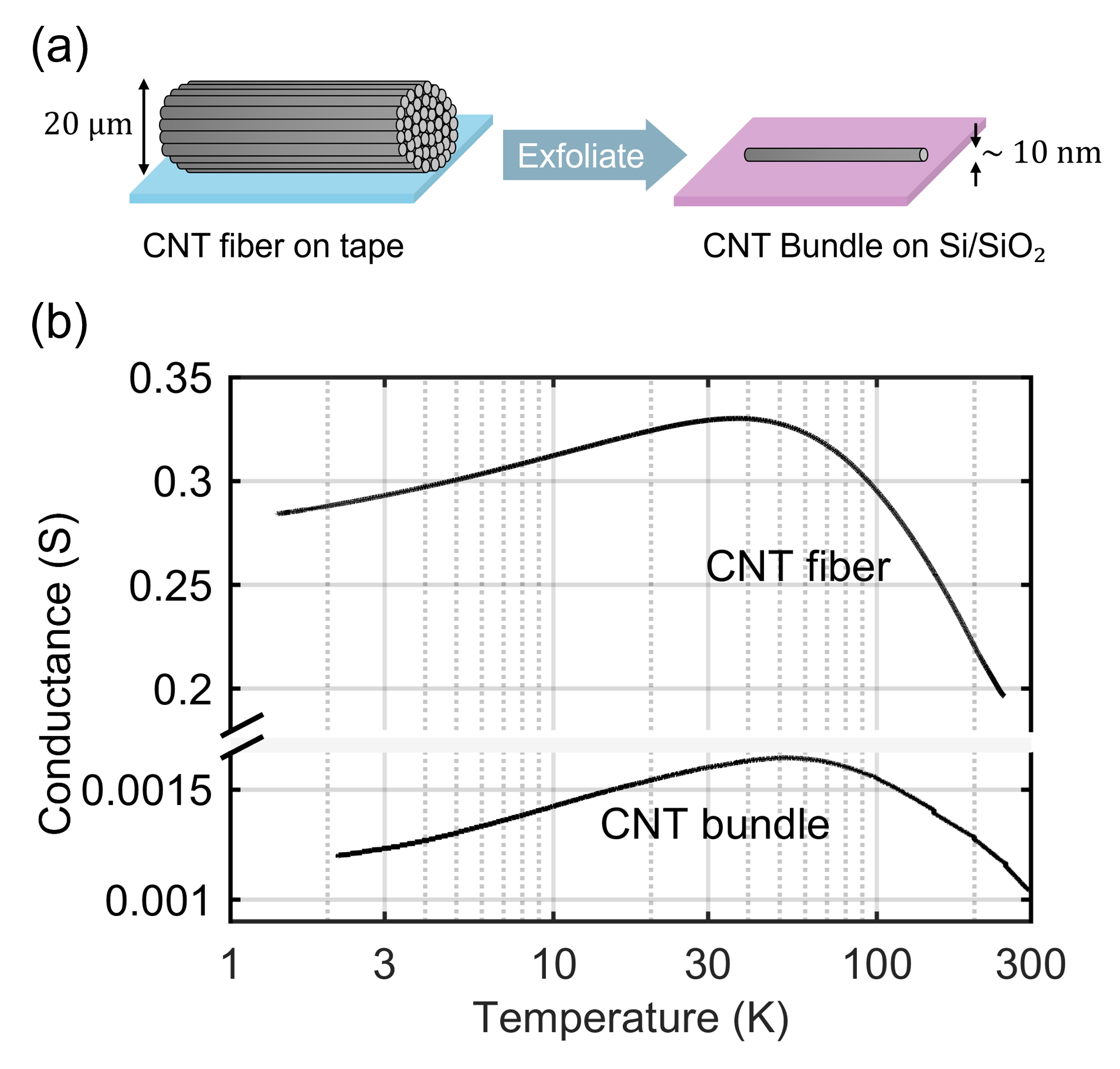}
  \caption{\label{fig:Fig1}
  (a) Schematic illustration of the exfoliation process from a macroscopic CNT fiber ($\sim$20\,$\upmu$m in diameter) to an individual CNT bundle ($\sim$10\,nm in diameter) on a Si/SiO$_2$ substrate.  
  A detailed characterization of the fiber’s hierarchical structure is reported in Ref.~\cite{Yu2025}.  
  (b) Temperature dependence of the four-terminal conductance for an 8-$\upmu$m CNT-bundle segment, compared with the CNT-fiber data reproduced from Ref.~\cite{Yu2025}.  
  Both samples show metallic behavior at high temperatures and a moderate upturn in resistance below $\sim$50\,K, consistent with quantum corrections arising from weak localization and Altshuler–Aronov interactions.
  }
\end{figure}

Figure~\ref{fig:Fig1}(a) schematically illustrates the exfoliation process from macroscopic CNT fibers to individual bundles used for device fabrication.  
The hierarchical architecture of the parent CNT fibers has been characterized extensively in Ref.~\cite{Yu2025}, which demonstrated their multiscale organization from individual nanotubes ($\sim$1.5\,nm diameter) forming tightly packed bundles (10–100\,nm) that align into macroscopic fibers ($\sim$20\,$\upmu$m diameter).  
This multilevel organization provides quasi-one-dimensional conduction pathways within bundles embedded in a three-dimensional network.  
All CNT bundles investigated here were exfoliated from fibers prepared under identical processing conditions.


Figure~\ref{fig:Fig1}(b) shows the temperature ($T$)-dependent four-terminal resistance of an 8-$\upmu$m CNT-bundle segment, which exhibits metallic behavior at high $T$ and a moderate upward trend with decreasing $T$ below $\sim$50\,K.  
This temperature dependence closely parallels that of macroscopic CNT fibers, with the fiber data reproduced from Ref.~\cite{Yu2025}, where the conductance likewise increases upon cooling before turning downward at low $T$.
In both systems, the low-$T$ conductance downturn is consistent with quantum corrections to the conductivity commonly associated with WL and electron--electron interaction effects, indicating a common diffusive transport regime across hierarchical length scales.
The bundle, exfoliated from the parent fiber, exhibits a somewhat stronger resistance upturn, possibly reflecting a slight dedoping effect introduced during transfer and electrode fabrication, where mild heating is unavoidable.


\begin{figure}[t]
  \centering
  \includegraphics[width= \linewidth]{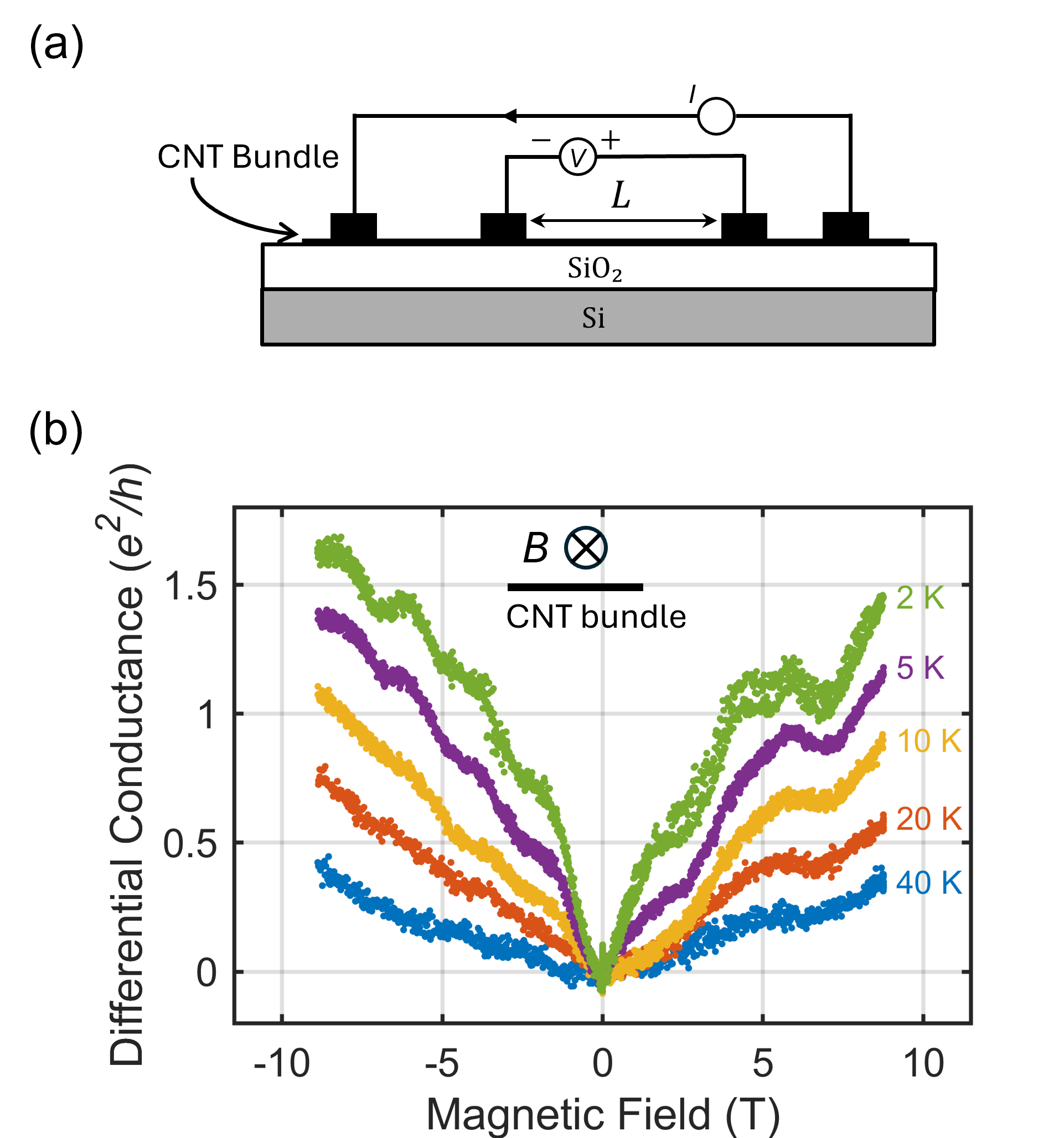}
  \caption{
    \label{fig:fig3_local}
    (a)~Schematic diagram of the local four-terminal measurement configuration used for CNT-bundle devices on Si/SiO$_2$ substrates.
    (b)~Magnetoconductance, expressed in units of $e^2/h =$ 3.874~$\times$~10$^{-5}$\,S, of an $L=8\,\upmu$m CNT-bundle segment measured at several temperatures for perpendicular magnetic-field orientations.
    A pronounced positive low-field magnetoconductance reflects the suppression of weak localization, while reproducible aperiodic fluctuations correspond to universal conductance fluctuations.
  }
\end{figure}

Figure~\ref{fig:fig3_local} shows the magnetoconductance of a representative 8-$\upmu$m CNT-bundle segment measured at various temperatures.  
The conductance exhibits a pronounced positive magnetoconductance centered around zero field, consistent with the suppression of WL by an applied magnetic flux.  
Superimposed on this background are reproducible aperiodic fluctuations as a function of $B$ in both perpendicular and parallel orientations.  
These fluctuations remain identical upon repeated field sweeps at a fixed $T$ but vary between cooldowns, characteristic of UCF; see Fig.~S1 for a direct comparison of up- and down-sweep data at $T=5$~K. 
A slight asymmetry with respect to field polarity is also observed. In a four-terminal measurement such as this, the lack of symmetry with field polarity is itself a signature of phase coherence along the length of the 8-$\upmu$m segment\cite{buttiker1986fieldsymm}.
Both the WL and UCF features diminish in amplitude as $T$ increases, yet remain discernible even at 20\,K, demonstrating robust phase-coherent transport over micrometer-scale distances.  
Similar behaviors were observed in other segments of different lengths, as discussed in the Supplemental Material.

Quantitative analysis of magnetoconductance presents inherent challenges for hierarchical CNT systems.  
Theoretical formulations of WL and UCF were developed for homogeneous materials, where the conversion between the phase-coherence time $\tau_{\phi}$ and the length $L_{\phi}$ requires knowledge of the electronic diffusion constant $D$.  
In the present case, the bundle conductivity reflects a complex network of intertube and intratube transport, which makes a single $D$ challenging to define.  
Moreover, the characteristic $B$ scale of the UCF is sufficiently large that, within experimentally accessible magnetic fields, the available $B$ range does not allow sufficient statistical averaging to rigorously determine the correlation field, $B_{\mathrm{c}}$.  
The appropriate functional form for fitting the WL magnetoconductance also depends on the effective dimensionality: in the quasi-1D limit ($t,w < L_{\phi}$) and the 3D limit ($t, w > L_{\phi}$), where $t$ and $w$ are the thickness and width of the conductor, respectively, distinct analytic expressions apply.  
For an approximately cylindrical bundle with diameter $d \sim L_{\phi}$, both descriptions can yield  fits of comparable quality, with quasi-1D analysis giving $L_{\phi} < d$ and 3D analysis giving $L_{\phi} > d$.  

\begin{figure}[t]
  \centering
  \includegraphics[width=\linewidth]{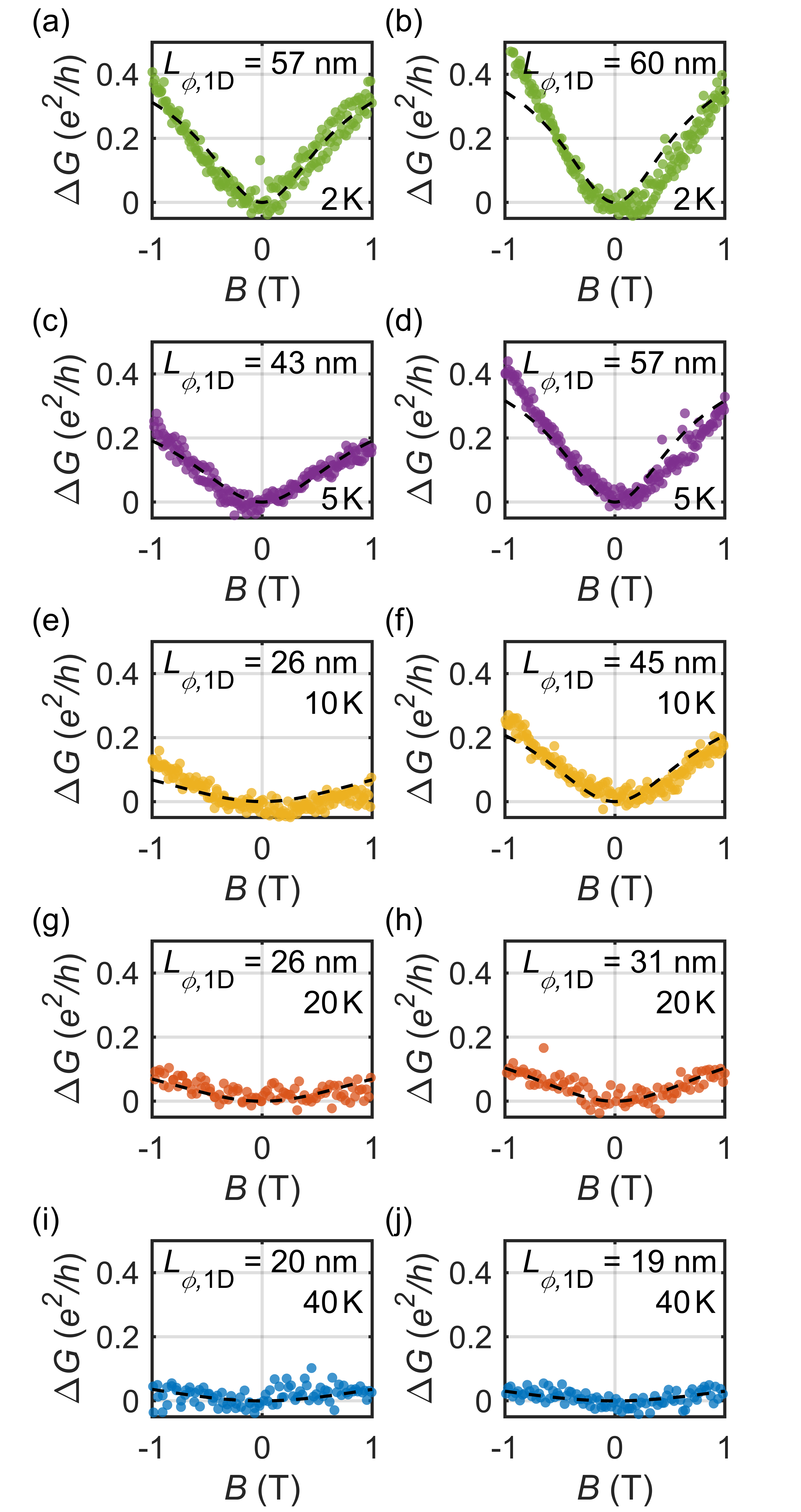}
  \caption{\label{fig:fig4}
  Magnetoconductance $\Delta G$ for the 8\,$\upmu$m CNT-bundle segment measured at different temperatures (2–40\,K), plotted as a function of magnetic field $B$.  
  Each panel shows the experimental data (dots) together with the quasi-1D weak localization fit (black dashed curve) using Eq.~(1).  
  The extracted phase-coherence length $L_{\phi,\mathrm{1D}}$ is indicated in each panel.  
  Panels (a), (c), (e), (g), and (i) correspond to measurements with the magnetic field \textit{parallel} to the bundle axis, while panels (b), (d), (f), (h), and (j) correspond to the field \textit{perpendicular} to the bundle axis.  
  }
\end{figure}

At low $T$, the 8-$\upmu$m segment exhibits a pronounced positive magnetoconductance around zero field (Fig.~\ref{fig:fig3_local}), consistent with suppression of WL by the $B$.  We note that, by inspection, the field scale associated with this WL magnetoconductance is comparable to the field scale associated with the UCF; this is not surprising given that both WL and field-dependent UCF result from threading of flux through coherent trajectories.
To quantify phase coherence, we analyzed the data using a quasi-1D WL model for diffusive conductors~\cite{altshuler1981, Beenakker1991, Yu2025},
\begin{align}
\Delta G(B) =
-\frac{2 e^2}{L h}
\raisebox{-2ex}{$\left(\vphantom{\dfrac{1}{\sqrt{L_{\phi,\mathrm{1D}}^{-2} + \tfrac{A_\mathrm{b}}{3 \ell_B^4}}}}\right.$}%
\frac{1}{\sqrt{L_{\phi,\mathrm{1D}}^{-2} + \dfrac{A_\mathrm{b}}{3 \ell_B^4}}}
- L_{\phi,\mathrm{1D}}
\raisebox{-2ex}{$\left.\vphantom{\dfrac{1}{\sqrt{L_{\phi,\mathrm{1D}}^{-2} + \tfrac{A_\mathrm{b}}{3 \ell_B^4}}}}\right)$},
\label{eq:1D_final}
\end{align}
where $e$ is the electronic charge, $h$ is the Planck constant, $L$ is the channel length, $A_\mathrm{b}$ is the bundle cross-sectional area, $\ell_B=(\hbar/eB)^{1/2}$ is the magnetic length, and $\hbar = h/(2\pi)$.  
Rather than assuming a particular decoherence mechanism, we treated $L_{\phi}$ as a free parameter at each $T$.
For both perpendicular and parallel $B$ orientations, $L_{\phi}$ increases from $\sim$25\,nm at 20\,K to $\sim$60\,nm at 2\,K (Fig.~\ref{fig:fig4}).  
This monotonic growth of $L_{\phi}$ with decreasing $T$ is consistent with quasi-1D Nyquist electron–electron dephasing at higher $T$, for which the dephasing rate follows $\tau_{\phi}^{-1}\propto T^{2/3}$ and hence $L_{\phi}\propto T^{-1/3}$, as predicted by the Altshuler–Aronov–Khmelnitskii theory for diffusive wires~\cite{altshuler1982}.  
Below $\sim$5\,K, $L_{\phi}$ exhibits a partial saturation, plausibly arising from $T$-independent channels such as magnetic-impurity scattering or residual electromagnetic noise once phonons are frozen out.  

Comparison between perpendicular and parallel $B$ orientations further provides insight into anisotropic diffusion within the bundle, in which individual CNTs are aligned and packed.  
Representative magnetoconductance data for both perpendicular and parallel magnetic field configurations are shown in Fig.~S2 of the Supplemental Material.  One can attempt to model anisotropic diffusion through two constants, $D_{\perp}$ and $D_{\parallel}$, for diffusion across and along the bundle direction, respectively~\cite{al1981anomalous, ALTSHULER19851}.
Despite the geometric factor associated with the diffusion-constant ratio $D_{\perp}/D_{\parallel}$, the two orientations yield comparable WL magnitudes, with the parallel-field case exhibiting an even stronger $B$ dependence.  
This weak anisotropy indicates that the WL corrections are comparable for perpendicular and parallel magnetic-field orientations. The microscopic origin of this behavior remains unclear, and may reflect the complex interplay between anisotropic diffusion, intertube coupling, and phase-breaking processes within the bundle.

Superimposed on the WL background are reproducible UCF, aperiodic conductance fluctuations as a function of $B$ (Fig.~\ref{fig:fig3_local}).  
At 2\,K, the root-mean-square amplitude of these reaches $\delta G \sim 0.1\,e^{2}/h$ (where $e^2/h =$ 3.874~$\times$~10$^{-5}$\,S) even for an 8-$\upmu$m-long channel.  
For a quasi-1D conductor with $L \gg L_{\phi}$, ensemble averaging predicts~\cite{Lee1985ucf}
\begin{align}
\delta G \approx \frac{e^{2}}{h}\!\left(\frac{L_{\phi}}{L}\right)^{3/2}.  
\end{align}
If we use the coherence length inferred from the WL magnetoconductance, with $L_{\phi}\approx 50$\,nm and $L = 8\,\upmu$m, the expected UCF amplitude is only $\sim$10$^{-3}(e^{2}/h)$, two orders of magnitude smaller than observed.  
Thermal averaging would further suppress the variance by $(E_\text{c}/k_\text{B}T)^{1/2}$, where $E_\text{c}= \hbar D/L_{\phi}^{2}$ is the correlation energy; even with typical metallic diffusion constants, this factor is $\sim$0.5 at 2\,K.  
The observation of $\delta G$ roughly $10^{2} \times$  larger than the ensemble-averaged expectation therefore indicates incomplete statistical averaging.  
Such enhanced, slowly varying fluctuation patterns suggest that the coherence length associated with longitudinal intrabundle transport is much longer than that inferred from the field dependence.   
Moreover, the persistence of these UCF-like features up to 20–40\,K points to unexpectedly robust quantum coherence extending over micrometer-scale paths.
Such anomalously large, slowly varying fluctuation patterns indicate that a small number of high-transmission, phase-correlated paths dominate conduction, effectively reducing ensemble averaging and amplifying UCF amplitudes by two orders of magnitude.

\begin{figure}[t]
  \centering
  \includegraphics[width= \linewidth]{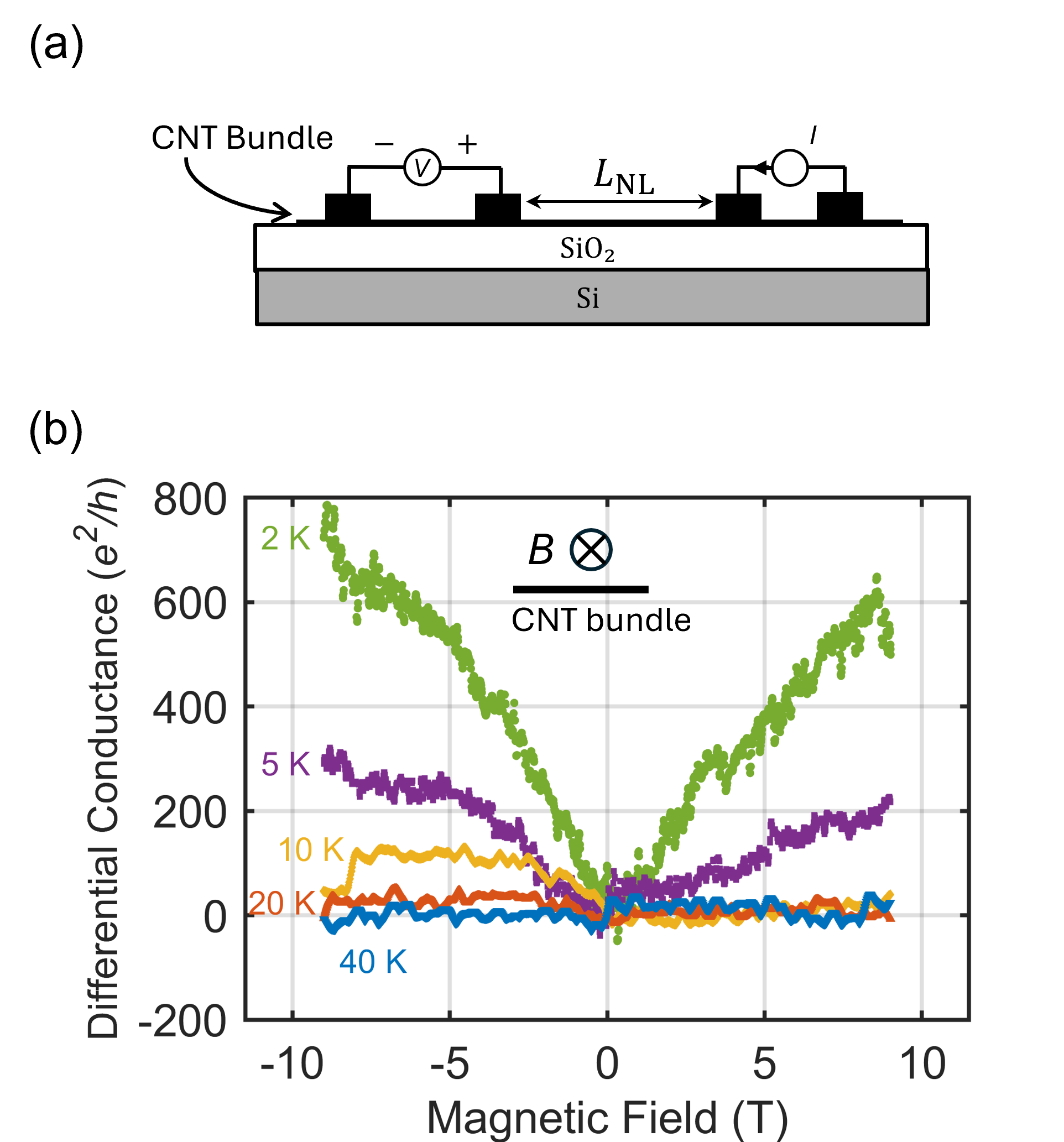}
  \caption{
    \label{fig:fig5_nonlocal}
    (a)~Schematic diagrams of the nonlocal measurement configurations used for CNT-bundle devices on Si/SiO$_2$ substrates.  
    (b)~Nonlocal magnetoconductance of a CNT-bundle device with $L_{\text{NL}}=4\,\upmu$m separation between current and voltage leads.  
    The positive low-field response at 2\,K diminishes with temperature, indicating micrometer-scale phase coherence across the bundle network.
  }
\end{figure}

%
The unexpectedly large UCF amplitude and its persistence to high $T$ imply that electronic transport is coherent over micrometer-scale distances, comparable to the lengths of the constituent CNTs.  
To directly test this long-range coherence, we fabricated nonlocal devices in which the current and voltage leads were separated by 4\,$\upmu$m (Fig.~\ref{fig:fig5_nonlocal}).  
At 2\,K, the nonlocal voltage exhibits a clear positive magnetoconductance that diminishes with increasing $T$, mirroring the WL–like response of the local channel.  
Because no current flows directly through the nonlocal segment, this signal cannot arise from classical diffusion, but instead indicates phase-coherent propagation of electronic wave functions across micrometer distances—far exceeding the $L_{\phi}$ extracted from local WL fits.  
In diffusive metals, nonlocal magnetoconductance vanishes when current and voltage leads are separated beyond $L_{\phi}$. In contrast, the persistence of a WL-like nonlocal signal across 4\,$\upmu$m in our bundles directly evidences long-range phase correlations.


These observations reveal a coexistence of phase-coherent phenomena occurring over distinct length scales within a single bundle.
While WL and the field scale of UCF indicate short diffusive trajectories that lose phase coherence over tens of nanometers, the large-amplitude fluctuations and nonlocal responses imply that certain electronic pathways preserve phase over much longer distances.
The microscopic origin of this coexistence remains unresolved, but it suggests that multiple dephasing channels or distinct transport modes may contribute to the observed behavior within the same mesoscopic conductor.

The apparent discrepancy between the phase-coherence length extracted from WL and the much longer coherence implied by the UCF amplitude and nonlocal transport highlights that quantum interference in CNT bundles manifests over multiple characteristic length scales. Different experimental probes emphasize distinct classes of electronic trajectories and therefore may not be characterized by a single, unified coherence length.

In particular, the magnetic field scale governing WL and the field-dependent UCF reflects the ability to thread applied flux through sets of trajectories between starting and ending points. In aligned CNT bundles, such sets of trajectories necessarily involve transverse motion between neighboring nanotubes, since the magnetic fields accessible here cannot thread an appreciable flux through the circumference of an individual CNT. If intertube motion requires inelastic
scattering processes—such as phonon-assisted or electron–electron interactions
with small energy transfer to satisfy momentum conservation—then trajectories
involving transverse diffusion are expected to experience enhanced dephasing.
As a result, coherence lengths inferred from magnetoconductance field scales may primarily reflect the limited phase coherence associated with intertube motion.

At the same time, phase coherence along individual nanotubes can persist over
much longer distances. Consequently, the length scales governing longitudinal
ensemble averaging, the magnitude of the UCF, and the presence of nonlocal signals may instead be set predominantly by intratube coherence, even when transverse
motion between tubes is partially dephased. Within this picture, short coherence lengths extracted from magnetic-field scales and much longer coherence implied by fluctuation amplitudes and nonlocal transport are not contradictory, but rather probe different aspects of transport in an anisotropic, heterogeneous conductor.

Similar phenomenology has been reported in other quasi-1D systems,
such as Bi$_4$Br$_4$, where coherence lengths inferred from magnetotransport
coexist with large UCF amplitudes and signatures of long-range phase-coherent
transport~\cite{lefeuvre2025quantum}. In that case, these observations were
reconciled by recognizing that different quantum-interference probes are
sensitive to distinct transport regions and classes of electronic trajectories.
Developing a quantitative theoretical framework that captures such coexistence
in CNT bundles will require models that go beyond homogeneous diffusive transport and explicitly incorporate anisotropic conduction, intertube coupling, and energy-dependent dephasing mechanisms.


In summary, we have performed low-temperature magnetotransport measurements on
individual CNT bundles exfoliated from highly conductive aligned fibers and
observed strikingly disparate coherence scales inferred from different quantum
interference probes.
WL magnetoconductance yields phase-coherence lengths of only tens of nanometers, while the large UCF amplitudes and the persistence of nonlocal magnetoconductance demonstrate phase-coherent transport over micrometer
length scales within the same bundle.

These results challenge the applicability of the conventional single-scale picture of mesoscopic coherence in homogeneous conductors and highlight the importance of trajectory- dependent quantum interference in heterogeneous, anisotropic systems.
Aligned CNT bundles thus provide a valuable experimental platform for exploring
how quantum coherence manifests in complex assemblies of coupled one-dimensional conductors.
A complete understanding of the observed coexistence of short- and long-range
coherence will require further theoretical developments that explicitly account
for anisotropic transport, intertube coupling, and energy-dependent dephasing.

\textit{Acknowledgments}--
We acknowledge the Shared Equipment Authority (SEA) at Rice University for providing access to instrumentation used in this work, and thank the SEA staff for their assistance with sample preparation and materials characterization. J.K.\ acknowledges support from the Robert A.\ Welch Foundation through Grant No.\ C-1509 and the Air Force Office of Scientific Research through Grant No.\ FA9550-22-1-0382.  M.P.\ acknowledges support from the Robert A.\ Welch Foundation through Grant No.\ C-1668.  D.N., R.L., and T.L.\ acknowledge support from NSF DMR-2102028.  G.W., Y.S., J.K., M.F., D.N., and M.P. acknowledge support from the Carbon Hub, a nonprofit institute which receives corporate funding from Shell, Mitsubishi Corporation (Americas), Prysmian, Saudi Aramco, Huntsman, ExxonMobil, Chevron, SABIC, and TotalEnergies.

\bibliographystyle{apsrev4-2}
\bibliography{apssamp} 

\end{document}


\title{Supplemental Material for\\
``Disparate Quantum Corrections to Conduction in Carbon Nanotube Bundles''}

\author{Shengjie Yu}
\affiliation{Department of Electrical and Computer Engineering, Rice University, Houston, Texas 77005, USA}
\affiliation{Applied Physics Graduate Program, Smalley-Curl Institute, Rice University, Houston, Texas 77005, USA}
\affiliation{Carbon Hub, Rice University, Houston, Texas 77005, USA}

\author{Zhengyi Lu}
\affiliation{Department of Electrical and Computer Engineering, Rice University, Houston, Texas 77005, USA}
\affiliation{Department of Physics and Astronomy, Rice University, Houston, Texas 77005, USA}

\author{Renjie Luo}
\affiliation{Department of Physics and Astronomy, Rice University, Houston, Texas 77005, USA}

\author{Tanner Legvold}
\thanks{Deceased}
\affiliation{Department of Physics and Astronomy, Rice University, Houston, Texas 77005, USA}

\author{Natsumi Komatsu}
\affiliation{Department of Electrical and Computer Engineering, Rice University, Houston, Texas 77005, USA}
\affiliation{Carbon Hub, Rice University, Houston, Texas 77005, USA}
\affiliation{Department of Bioengineering, University of Illinois Urbana–Champaign, Urbana, Illinois 61801, USA}

\author{Liyang~Chen}
\affiliation{Applied Physics Graduate Program, Smalley-Curl Institute, Rice University, Houston, Texas 77005, USA}
\affiliation{Department of Physics and Astronomy, Rice University, Houston, Texas 77005, USA}

\author{Oliver S.\ Dewey}
\affiliation{Carbon Hub, Rice University, Houston, Texas 77005, USA}
\affiliation{Department of Chemical and Biomolecular Engineering, Rice University, Houston, Texas 77005, USA}

\author{Lauren W.\ Taylor}
\affiliation{Carbon Hub, Rice University, Houston, Texas 77005, USA}
\affiliation{Department of Chemical and Biomolecular Engineering, Rice University, Houston, Texas 77005, USA}
\affiliation{Department of Chemical and Biomolecular Engineering, The Ohio State University, Columbus, OH 43210}

\author{Huaijin Sun}
\affiliation{Department of Physics and Astronomy, Rice University, Houston, Texas 77005, USA}

\author{Matteo Pasquali}
\affiliation{Carbon Hub, Rice University, Houston, Texas 77005, USA}
\affiliation{Department of Chemical and Biomolecular Engineering, Rice University, Houston, Texas 77005, USA}
\affiliation{Smalley--Curl Institute, Rice University, Houston, Texas 77005, USA}
\affiliation{Department of Materials Science and NanoEngineering, Rice University, Houston, Texas 77005, USA}
\affiliation{Department of Chemistry, Rice University, Houston, Texas 77005, USA}

\author{Geoff Wehmeyer}
\affiliation{Carbon Hub, Rice University, Houston, Texas 77005, USA}
\affiliation{Department of Mechanical Engineering, Rice University, Houston, Texas 77005, USA}
\affiliation{Smalley--Curl Institute, Rice University, Houston, Texas 77005, USA}

\author{Matthew S.\ Foster}
\affiliation{Carbon Hub, Rice University, Houston, Texas 77005, USA}
\affiliation{Department of Physics and Astronomy, Rice University, Houston, Texas 77005, USA}
\affiliation{Smalley--Curl Institute, Rice University, Houston, Texas 77005, USA}

\author{Junichiro Kono}
\thanks{Corresponding author: kono@rice.edu}
\affiliation{Department of Electrical and Computer Engineering, Rice University, Houston, Texas 77005, USA}
\affiliation{Carbon Hub, Rice University, Houston, Texas 77005, USA}
\affiliation{Department of Physics and Astronomy, Rice University, Houston, Texas 77005, USA}
\affiliation{Smalley--Curl Institute, Rice University, Houston, Texas 77005, USA}
\affiliation{Department of Materials Science and NanoEngineering, Rice University, Houston, Texas 77005, USA}

\author{Douglas Natelson}
\thanks{Corresponding author: natelson@rice.edu}
\affiliation{Department of Electrical and Computer Engineering, Rice University, Houston, Texas 77005, USA}
\affiliation{Carbon Hub, Rice University, Houston, Texas 77005, USA}
\affiliation{Department of Physics and Astronomy, Rice University, Houston, Texas 77005, USA}
\affiliation{Smalley--Curl Institute, Rice University, Houston, Texas 77005, USA} 
\affiliation{Department of Materials Science and NanoEngineering, Rice University, Houston, Texas 77005, USA}

\date{\today}

\maketitle



\section{Reproducibility of Universal Conductance Fluctuations}
\begin{figure}[t]
  \centering
  \includegraphics[width=0.95\linewidth]{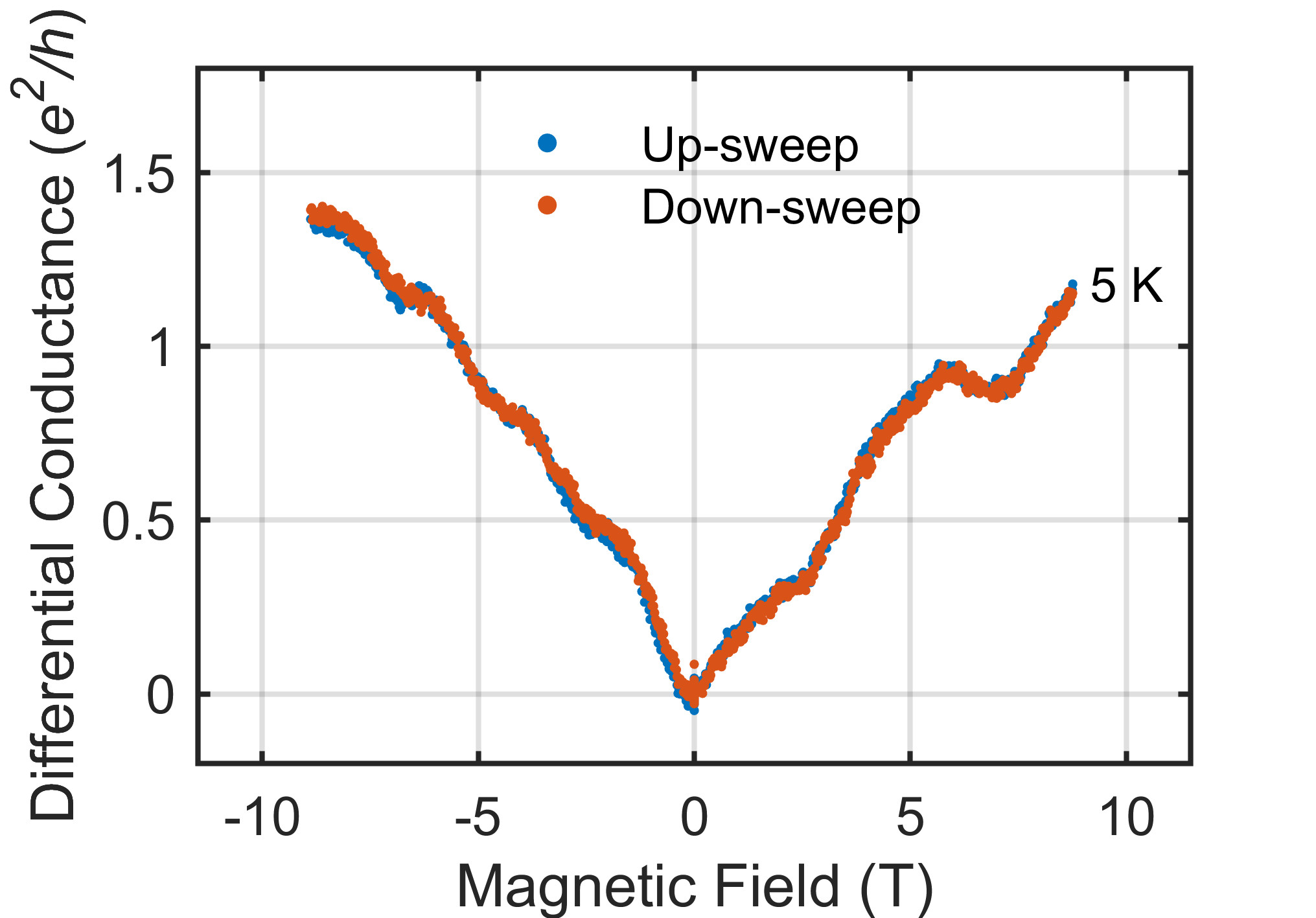}
  \caption{\label{fig:SI_updown}
Magnetoconductance traces measured at $T=5$~K under a perpendicular magnetic field
for increasing field $B$ (up-sweep) and decreasing $B$ (down-sweep).
The down-sweep trace is vertically offset for clarity.
The reproducibility of the aperiodic fluctuation pattern supports its identification
as UCF.}
\end{figure}

To verify the reproducibility of the aperiodic magnetoconductance fluctuations
under perpendicular magnetic fields,
we compare measurements taken during increasing- and decreasing-field sweeps
at a fixed temperature.
Figure~S\ref{fig:SI_updown} shows representative data at $T=5$~K.
The two traces exhibit the same fluctuation pattern, confirming their identification
as universal conductance fluctuations (UCF).

\section{Magnetoconductance of CNT-bundle segments with different lengths}

To examine the reproducibility of the magnetoconductance features discussed in the main text, we present in Fig.~S2 representative data measured on CNT-bundle segments with channel
lengths of $L=8~\upmu$m and $L=4~\upmu$m. 
The data for the $L=8~\upmu$m segment shown in Fig.~S2(a) are the same as those presented in Fig.~2(b) of the main text. 
For each length, measurements were performed over the same temperature range and for both
perpendicular and parallel magnetic-field orientations.

For both channel lengths, a pronounced positive magnetoconductance centered around zero field is observed at low temperatures, consistent with the suppression of weak localization.
Superimposed on this background are reproducible, aperiodic conductance fluctuations,
whose characteristic field scale and temperature dependence are consistent with
universal conductance fluctuations.
In all cases, the amplitude of both WL and UCF features decreases systematically with
increasing temperature, while remaining discernible up to $\sim$20~K.

While the detailed fluctuation patterns differ between segments and between lengths,
reflecting the mesoscopic sensitivity to disorder configuration,
the overall behavior is robust across device length.
These results demonstrate that the coexistence of WL and UCF, as well as their temperature
evolution, is an intrinsic feature of phase-coherent transport in CNT bundles over
micrometer-scale distances.

\begin{figure*}[t]
  \centering
  \includegraphics[width=\textwidth]{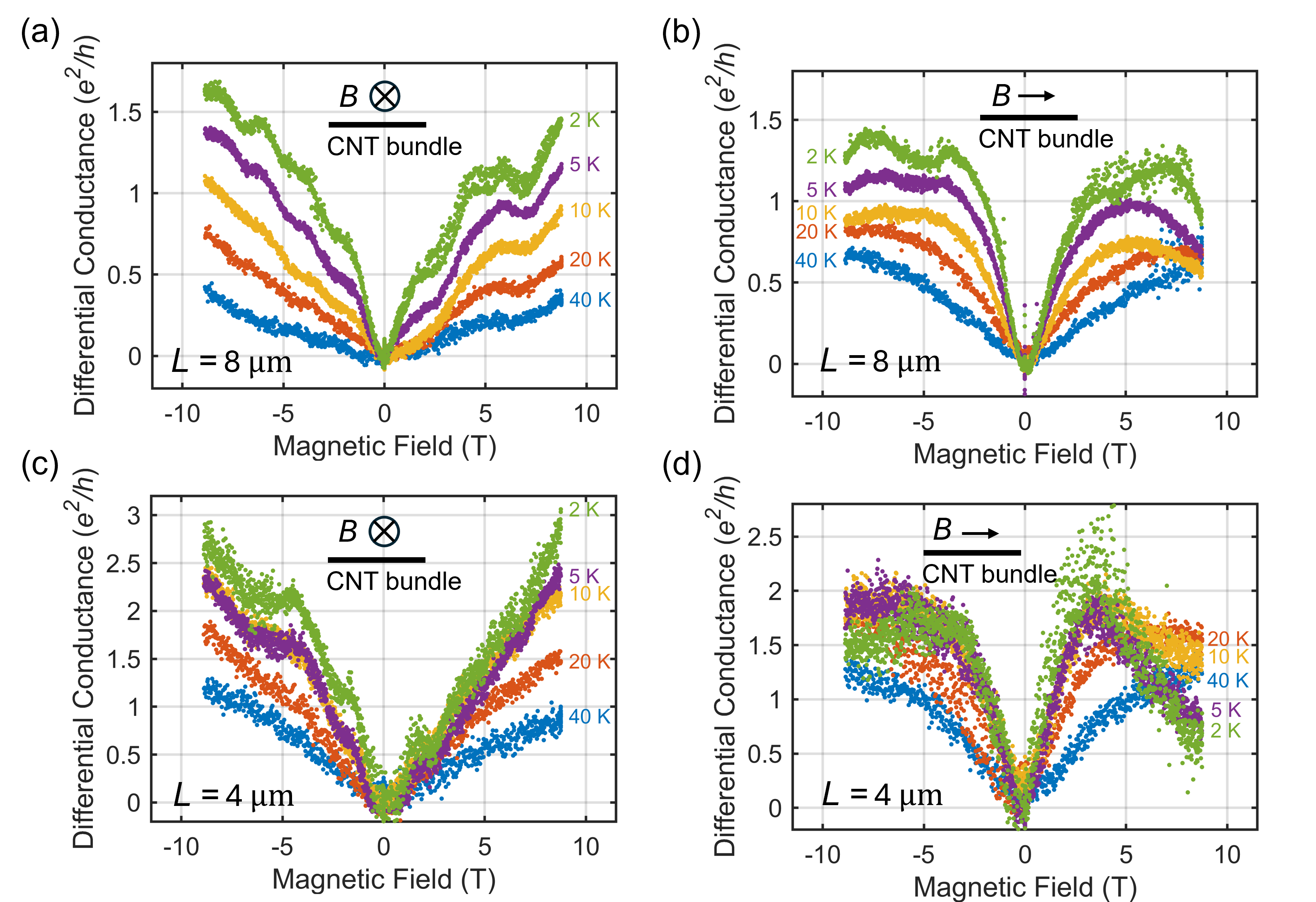}
  \caption{\label{fig:SI_S2}
  Magnetoconductance of representative CNT-bundle segments with channel lengths
  of $L=8~\upmu$m (top row) and $L=4~\upmu$m (bottom row), measured at various temperatures.
  For each length, data are shown for perpendicular ($B \perp$ bundle axis, left column)
  and parallel ($B \parallel$ bundle axis, right column) magnetic-field orientations.
  In all cases, a pronounced positive magnetoconductance centered around zero field is observed,
  consistent with the suppression of weak localization.
  Superimposed on this background are reproducible aperiodic conductance fluctuations,
  characteristic of universal conductance fluctuations.
  The amplitude of both WL and UCF features decreases with increasing temperature,
  while remaining discernible up to $\sim$20~K.
  }
\end{figure*}

\bibliographystyle{apsrev4-2}